\documentclass{article}

\usepackage{arxiv}

\usepackage[utf8]{inputenc} 
\usepackage[T1]{fontenc}    
\usepackage{hyperref}       
\usepackage{url}            
\usepackage{booktabs}       
\usepackage{amsfonts}       
\usepackage{nicefrac}       
\usepackage{microtype}      
\usepackage{lipsum}         
\usepackage{graphicx}
\usepackage{doi}

\usepackage{amsmath}
\usepackage{amssymb}
\usepackage{pifont}
\usepackage{paralist}
\usepackage{subcaption}
\usepackage{array,multirow}
\usepackage[table]{xcolor}
\newcolumntype{P}[1]{>{\centering\arraybackslash}p{#1}}

\title{Deep Reinforcement Learning for Delay-Optimized Task Offloading in Vehicular Fog Computing}

\newif\ifuniqueAffiliation
\uniqueAffiliationtrue

\author{ \href{}{\includegraphics[scale=0.00]{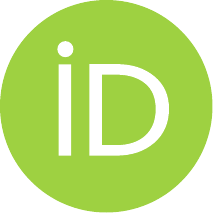}\hspace{1mm}Mohammad Parsa Toopchinezhad} \\
	Computer Engineering and \\Information Technology Department\\
	Razi University\\
	Kermanshah, Iran\\
	\texttt{ptoopchinzhad@gmail.com}\\
	\And
	\href{}{\includegraphics[scale=0.00]{orcid.pdf}\hspace{1mm}Mahmood Ahmadi} \\
	Computer Engineering and \\Information Technology Department\\
	Razi University\\
	Kermanshah, Iran\\
	\texttt{m.ahmadi@razi.ac.ir}\\
}

\renewcommand{\shorttitle}{Machine Learning Approaches for Active Queue Management}

\hypersetup{
pdftitle={Deep Reinforcement Learning for Delay-Optimized Task Offloading in Vehicular Fog Computing},
pdfsubject={Task Offloading, ML, RL},
pdfauthor={Mohammad Parsa Toopchinezhad, Mahmood Ahmadi},
pdfkeywords={Task Offloading, ML, RL},
}

\begin{document}
\maketitle

\begin{abstract}
The imminent rise of autonomous vehicles (AVs) is revolutionizing the future of transport. The Vehicular Fog Computing (VFC) paradigm has emerged to alleviate the load of compute-intensive and delay-sensitive AV programs via task offloading to nearby vehicles. Effective VFC requires an intelligent and dynamic offloading algorithm. As a result, this paper adapts Deep Reinforcement Learning (DRL) for VFC offloading. First, a simulation environment utilizing realistic hardware and task specifications, in addition to a novel vehicular movement model based on grid-planned cities, is created. Afterward, a DRL-based algorithm is trained and tested on the environment with the goal of minimizing global task delay. The DRL model displays impressive results, outperforming other greedy and conventional methods. The findings further demonstrate the effectiveness of the DRL model in minimizing queue congestion, especially when compared to traditional cloud computing methods that struggle to handle the demands of a large fleet of vehicles. This is corroborated by queuing theory, highlighting the self-scalability of the VFC-based DRL approach. \end{abstract}


\section{Introduction}

Born from the ever-expanding growth of smart devices, the Internet of Vehicles (IoV) has emerged as a promising paradigm for enabling safe, autonomous and more pleasant driving experiences \cite{behravan2023comprehensive}. Within the IoV model, vehicles exchange and analyze a wide range of data to enhance traffic control, optimize parking efficiency and promote vehicle platooning, while simultaneously bolstering safety through the provision of hazard warnings, collision detection, and accident reporting. Furthermore, IoV applications can facilitate entertainment services such as in-car multimedia and gaming programs, offering passengers a more enjoyable and interactive time while on the road \cite{hamdi2022task}.

Undeniably, this plethora of applications requires a significant amount of computational power, which may prove unfeasible to execute locally on a single vehicle. At first glance, utilizing the immense hardware resources of the cloud appears as an appealing solution. However, the typically distant geographical locations of these datacenters introduce unpredictable latency which falls short of the sensitive, real-time nature of most IoV applications \cite{mseddi2023centralized}.

To combat these constraints, researchers have proposed Vehicular Fog Computing (VFC) as an innovative approach for IoV task offloading \cite{behravan2023comprehensive}. Inspired by fog computing, VFC employs the resources of nearby vehicles (parked or moving), Road Side Units (RSU) and Base Stations (BS) to process vehicular emitted tasks. This paradigm of task offloading has been shown to greatly reduce task response time, energy usage and monetary costs \cite{hamdi2022task}.

Another incentive for VFC adaptation is unlocking the huge pool of previously underutilized mobile computing devices. There is an estimated 1.5 billion cars globally, this number is projected to reach 2 billion by 2040 \cite{wefcars}. Assuming an average of 10 GFlops per vehicle, 1.5 billion cars can theoretically provide a computing capacity of 15000 Peta Flops. This number dwarfs even the most powerful supercomputer system of 2024 with 1195 Peta Flops \cite{top500}. Moreover, the advancement of wireless technologies such as 5G, 6G, and LTE provides a foundation for the high-bandwidth, low-latency communication necessary to fully realize the capabilities of VFC.

Despite the significant advantages, VFC introduces many complications that must be addressed, the main point, determining which node a particular task should be offloaded to. This issue is further exacerbated by the dynamic properties of vehicular networks, where vehicles are constantly moving and thus forming varying communication links, or disconnecting altogether.

Offloading algorithms are designed to optimize delay and energy consumption. These algorithms exhibit a variety of different forms, including exact, heuristic, and metaheuristic approaches \cite{hamdi2022task}. An increasingly favored strategy is applying Reinforcement Learning (RL) to the problem of task offloading. RL, a subset of Machine Learning (ML), is centered on instructing intelligent agents on problem-solving through repeated experimentation and learning. RL algorithms have shown efficiency in diverse domains such as autonomous driving, recommendation systems, and computer network optimization \cite{arulkumaran2017deep}.

This paper investigates a deep RL approach for delay minimization in VFC task offloading. To achieve this end, a novel movement model is applied for realistic vehicular movement based on grid planned cities. Furthermore, we openly release our implementation\footnote{https://github.com/Procedurally-Generated-Human/VFC-Offloading-RL}, providing a foundational simulation tool for future contributions to build upon. To the authors' knowledge, this consists as the first open-sourced RL environment for VFC offloading. The main contributions of the paper include:

\begin{itemize}
 \item Formulating the VFC offloading problem as a queuing network while utilizing a novel movement model.
 \item Implementation of an open-sourced, realistic RL environment for VFC offloading.
 \item Training a dynamic DRL model that outperforms traditional offloading techniques, yielding reduced delay and minimized queue congestion.
\end{itemize}

\section{Related Works}
In this section, we review previous research on both non-RL and RL methods of VFC task offloading.
\subsection{Task Offloading in VFC}
\begin{figure}[!t]
\centering
\includegraphics[width=10cm]{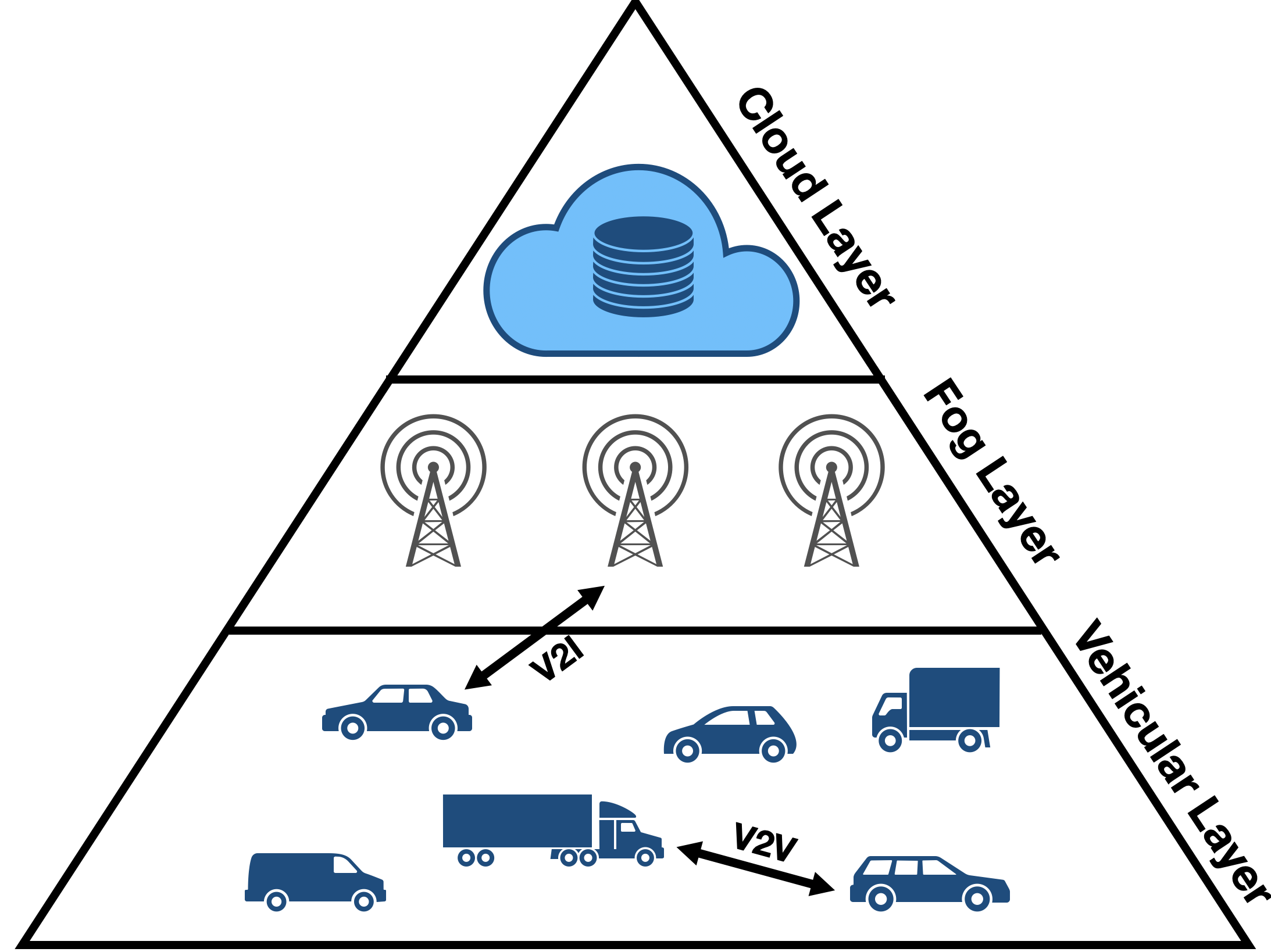}
\caption{VFC architecture; showcasing the three main layers along with V2I and V2V connections.}
\label{fig_sim}
\end{figure}

Task offloading is the process of data transfer from a resource-limited device to a nearby resource-rich one. Typically, an offloading process consists of three collaborators, namely the requester node, the communication link used for data transfer, and the service node that executes the offloaded task and returns the computed results. These communications transpire either between adjacent vehicles (V2V) or roadside infrastructure (V2I) (Figure \ref{fig_sim}). 

In a dynamic setting such as VFC, task offloading is a non-deterministic NP-hard problem \cite{zhu2018folo}, thus rendering exact solutions impractical. Therefore, many researchers have turned to approximate methods. Zhu et al. proposed an event-triggered task allocation strategy with joint linear programming and particle-swarm based optimization \cite{zhu2018folo}. Li et al. developed a contract-based scheme exploiting parked vehicles, which was optimized via a Lagrangian multiplier method \cite{li2018parked}. Wu et al. integrated predicted mobility information into the optimization problem \cite{wu2023joint}.

\subsection{RL-Based Task Offloading in VFC}
In RL, an agent interacts with its world by first observing the environment's state, taking a reasonable action and finally, receiving a reward based on its performance. This intuitive, data-driven method has been adapted for intelligent decision-making in many domains, in part, because of its model-free and applicable nature. Additionally, researchers have integrated deep neural networks into the conventional RL framework, birthing Deep Reinforcement Learning (DRL), further increasing the robust problem-solving power of RL \cite{arulkumaran2017deep}.

DRL models are prevalent in VFC and can be broadly categorized into two groups: Fog-Level Agent (FLA) and Edge-Level Agent (ELA) architectures (Figure 2). In FLA systems, the decision-making entities are located at the fog level, e.g., they reside in one or more RSU/BSs. These nodes can observe all vehicles within their range, and thus have access to a large section of the system's state. In \cite{mseddi2023centralized}, Mseddi et al. explored both a centralized and collaborative FLA approach, where RSU-based DRL agents govern how network resources should be allocated. Jamal et al. \cite{jamil2023irats}, applied Proximal Policy Optimization (PPO) to a similar FLA scheme.

In ELA systems, decision-making is devoted to the edge nodes, e.g., the mobile vehicles that produce tasks for offloading. Thus, each vehicle operates as a separate agent in an overall multi-agent environment. Clearly, in this setting, each agent only observes a limited number of neighboring vehicles and therefore must take actions without seeing the entire picture. On the positive side, ELA systems allow for distributed learning and help relieve a single node from making many offloading decisions. \cite{hou2023hierarchical} is an example of an ELA architecture where hierarchical Multi-Agent Reinforcement Learning (MARL) is used for training the agents. Wei et al. introduced an ELA approach for enhancing fog resource utilization via V2V-based trading and reported impressive results \cite{wei2022dynamic}.

A noticeable portion of prior VFC offloading research lacks realistic task specifications, hardware configurations, and accurate mobility models. Additionally, many of these works do not openly share their implementations, hindering the ability to build upon their findings. In contrast, this paper addresses both of these shortcomings by focusing on accurately depicting these aspects. Our work also emphasizes formulating the problem of VFC offloading as a queuing network, which is a more intrinsic method for simulating this kind of system.

\begin{figure}[htbp]
\centering
\subfloat[]{\includegraphics[height=5cm]{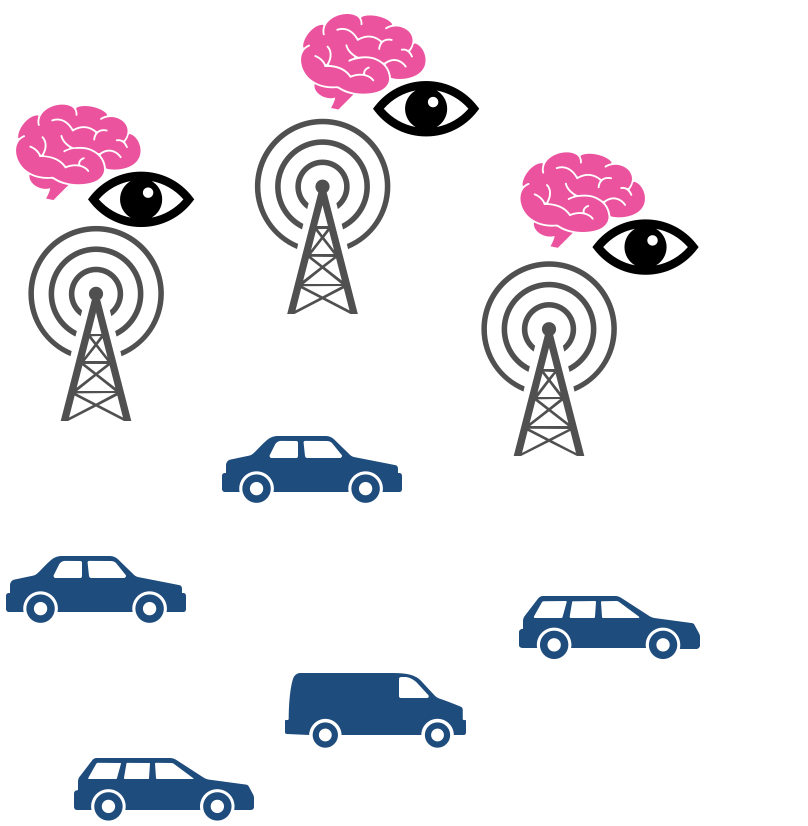}}
\subfloat[]{\includegraphics[height=5cm]{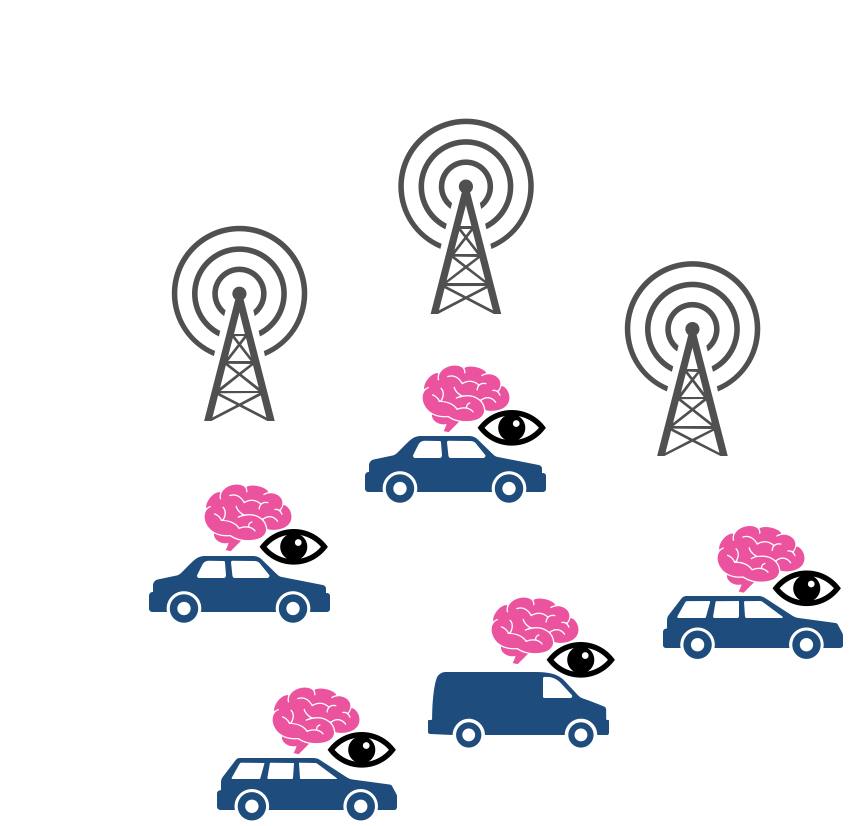}}
\caption{Two types of RL-based system architectures in VFC; fog-level agents (a) and edge-level agents (b).}
\end{figure}

\section{System Model}
A FLA model is considered where a single RSU is placed at the center of an $l$ by $w$ meter region in which $n$ clients and $m$ service vehicles reside. This region is modeled after a typical grid-planned city, where streets run at right angels, forming a symmetrical grid of same-sized city blocks (e.g., downtowns of major U.S. cities) (Figure 3).

Client vehicles circulating in this area send offloading requests to the RSU when faced with tasks beyond their local processing capability. The RSU then decides to process a task locally, delegate it to an appropriate service vehicle, or forward it to the cloud. In this setup, the primary objective of the allocation algorithm is to minimize average system delay. This section thoroughly outlines the simulation model, which has been deconstructed into several distinct components. A list of used notation exists at Table I.

\begin{figure}[htbp]
\centering
\subfloat[]{\includegraphics[height=5cm]{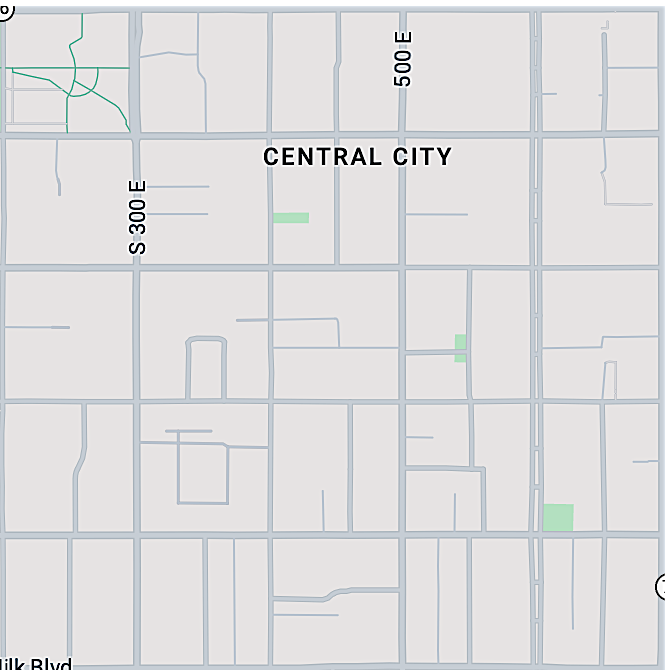}}
\subfloat[]{\includegraphics[height=5cm]{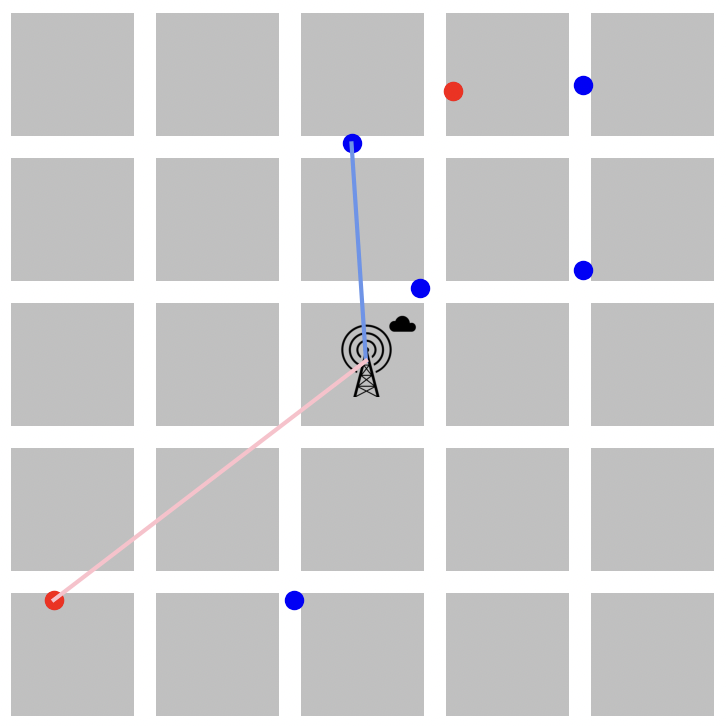}}
\caption{Downtown of Salt Lake City, U.S., showcasing its grid-like structure of 200 by 200 meter city blocks (a) (Map data @2024 Google). Simulated environment based on Salt Lake City blocks (b).}
\end{figure}

\subsection{Vehicle Model}
Each node plays a role as either a client or service vehicle. Client vehicles are mobile nodes that lack sufficient computational power and seek to offload tasks to parked service vehicles. Each service vehicle is characterized by a tuple, $V_j$: 
\begin{equation}\label{vehicle_service}
V_j = \{CPU_j, X_j, Y_j\}
\end{equation}
where the vehicle's computational capability (in MIPS) is denoted by $CPU_j$. $X_j$ and $Y_j$ represent the coordinates of the parked vehicle. 

Client vehicles are specified by the following four-tuple: 
\begin{equation}\label{vehicle_client}
V_i = \{\lambda_i, vel_i, X_i, Y_i\}
\end{equation}
where $\lambda_i$ designates the vehicle's average offloading request rate (tasks per second), while $vel_i$ determines its velocity. At the beginning of the simulation, each client vehicle is placed at a random legal position (i.e., within a road). It then commences movement in an arbitrary direction, maintaining it until reaching an intersection, where another random direction is re-selected. \par

\begin{table}[!t]
\renewcommand{\arraystretch}{1.3}
\caption{Notation used for system model}
\label{notation}
\centering
\begin{tabular}{|c||c|}
    \hline
	Symbol & Meaning\\
	\hline
	$l, w$ & Simulation area length \& width (m)\\
	$n$ & Number of client vehicles\\
	$m$ & Number of service vehicles\\
	$CPU_j$ & Processing power of vehicle $j$ (MIPS)\\
	$\lambda_i$ & Average task emission rate by client $i$\\
	$vel_i$ & Velocity of client $i$\\
	 $CU_x$ & Number of million instructions of task $x$ \\
	 $SZ_x$ & Size of task $x$ (Mb)\\
	 $TR_w$ & Transmission rate of medium $w$ (Mbps)\\
	 $B$ & Wireless channel bandwidth (MHz)\\
	 $N_0$ & Noise Power (dBm/Hz)\\
	 $P_T$ & Transmitter's transmission Power (W)\\
	 $G_T/G_R$ & Transmitting/Receiving directional gain (dBi)\\
    \hline
\end{tabular}
\end{table}      

\subsection{Task Model}
Each task is represented by a three-element tuple, $T_x$:
\begin{equation}\label{task}
T_x = \{CU_x, SZ_x, V_i\}
\end{equation} 
where $CU_x$ signifies the instruction count and $SZ_x$ indicates the size (in bits) of task $x$ generated by client $V_i$.

\begin{figure*}[!t]
\centering
\includegraphics[width=\columnwidth]{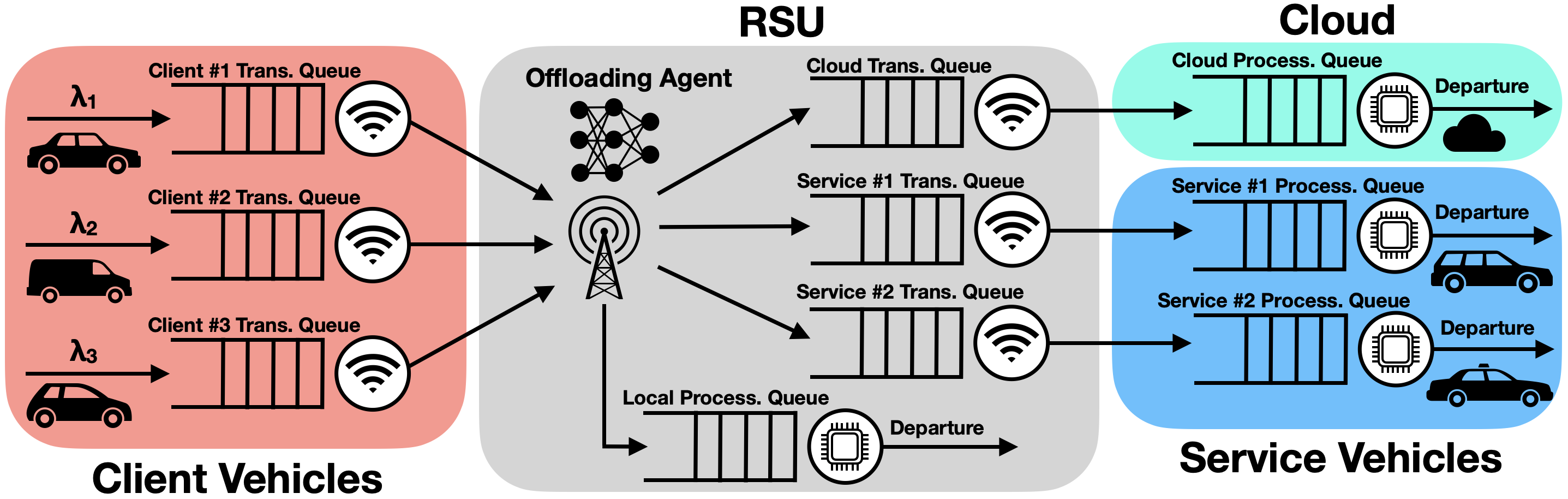}
\caption{VFC modeled as a queuing network; showcasing the possible paths an offloaded task can take to finish processing.}
\label{fig_sidsm}
\end{figure*}

\subsection{Delay Model}
The offloading system is modeled as a queuing network (Figure 4), where all queues operate in a FIFO fashion. The total delay experienced by a task consists of the entire time it spent inside the queuing network, which can be broken up as:
\begin{equation}\label{total_delay}
	d_\mathrm{total} = d_\mathrm{client}  + d_\mathrm{rsu}  + d_\mathrm{service} 
\end{equation} 
where $d_\mathrm{client}$, $d_\mathrm{rsu}$ and $d_\mathrm{service}$ specify the delay experienced at the client, RSU and service nodes, respectively. These values are calculated as follows:\\
\begin{equation}\label{client_delay}
d_\mathrm{client} = d_\mathrm{client-queue}  + d_\mathrm{client-trans}
\end{equation} 
\begin{equation}\label{rsu_delay}
d_\mathrm{rsu} = d_\mathrm{rsu-queue}  + d_\mathrm{rsu-trans}
\end{equation} 
\begin{equation}\label{service_delay}
d_\mathrm{service} = d_\mathrm{service-queue}  + d_\mathrm{service-process}
\end{equation} 

As an offloading task is created, it enters the client vehicle's transmission queue, where it must wait its turn to be transmitted to the RSU. The sum of this transmission delay ($d_\mathrm{client-trans}$), as well as any possible queuing delay ($d_\mathrm{client-queue}$), constitutes $d_\mathrm{client}$. Following this, the task arrives at the RSU, where it is placed at an appropriate out-going transmission queue as determined by the offloading algorithm. Once again, it experiences another set of queuing and transmission delays, summing up to $d_\mathrm{rsu}$. At this point, the task has reached its ultimate destination: the service node's execution queue. Here, it will wait for both computational resources to become available ($d_\mathrm{service-queue} $) and for its processing to be completed ($d_\mathrm{service-process}$), which together make up the experienced delay at the service node ($d_\mathrm{service}$).\par
$d_\mathrm{service-process}$ represents the amount of required time for task $x$ with size $CU_x$ (number of million instructions) to finish execution on the service node's hardware, with processor speed of $CPU_\mathrm{service}$ (million instructions per second):
\begin{equation}\label{process}
d_\mathrm{service-process} = \frac{CU_x}{CPU_\mathrm{service}}
\end{equation} 

$d_\mathrm{w-trans}$ specifies the time taken for transmitting task $x$ with size $SZ_x$ (bits) across medium $w$ with transfer rate $TR$ (bits/sec):
\begin{equation}\label{transmission-time}
d_\mathrm{w-trans} = \frac{CU_x}{TR_w}
\end{equation} 
when $w$ is a wireless medium, $TR_w$ it is calculated via Shanon's formula:
\begin{equation}\label{shanon}
TR = B.\log_2{(1+SNR)}
\end{equation} 
where $B$ (Hz) and $SNR$ are the channel bandwidth and signal-to-noise ratio, respectively. Assuming negligible receiver amplifier noise, $SNR$ is modeled as:
\begin{equation}\label{snr}
SNR = \frac{P_R}{N_0.W}
\end{equation} 
where $P_R$ denotes the reciver's signal and $N_0$ is the power of Additive White Gaussian Noise (AWGN). Considering a freespace propagation loss model, $P_R$ is equal to:
\begin{equation}\label{free-space-prop}
P_R = P_T.G_R.G_T. (\frac{\lambda}{4\pi d})^2
\end{equation} 
where $P_T$ is the transmitted power, $G_R$ and $G_T$ are "directional gains" of receiving and transmitting antennas respectively, $\lambda$ is the signal's wavelength and $d$ is the distance between transmitting and receiving nodes.

\section{Proposed Method}
Within our model, the core focus is on diminishing average task delay. To achieve this goal, a Markov Decision Process (MDP) is adopted as the mathematical framework to model the problem. MDPs are essential in RL for defining the environment. An MDP represents the environment as a tuple of $\langle S, A, T, R, \gamma \rangle$, where $S$ is the set of all states, $A$ is the set of possible actions, $T$ maps a state-action pair to a probability distribution of the next state, $R$ represents the reward function which determines the reward for taking an action in a specific state and $\gamma$ is a discount factor that determines the importance of future rewards relative to immediate ones. The objective of RL algorithms is to discover an optimal mapping of states to actions (known as a policy) that maximizes the overall discounted reward received over time \cite{arulkumaran2017deep}.

\subsection{State Space}
The state space contains the set of all possible states the agent can be in during a simulation. We posit a centralized decision-making agent located at the RSU. At the moment an offloading request ($T_x$) is received, the agent makes the following observation:
\begin{equation}\label{state}
s = \{SZ_x, CU_x, PP, DT, Q_t, Q_p\}
\end{equation} 
\begin{equation}\label{CPU}
PP = \{CPU_1, CPU_2, ...,  CPU_m, CPU_\mathrm{rsu}, CPU_\mathrm{cloud}\}
\end{equation}
\begin{equation}\label{dist}
DT = \{dist(V_1), dist(V_2), ..., dist(V_m) \}
\end{equation} 
where $SZ_x$ and $CU_x$ hold the details of the requested task, $PP$ and $DT$ are vectors containing the processing power and Euclidean distance of available service nodes relative to the RSU, respectively. $Q_t$ and $Q_p$ are also vectors containing the load of tasks currently waiting at each transmission and processing queue, respectively.
\subsection{Action Space}
After observing the current state, the agent must make an offloading decision, notated as $a_t$. Each task can be allocated to one of $M$ service vehicles ($a_t \in V$), forwarded to the cloud ($a_t = a_\mathrm{cloud}$) or alternatively processed locally by the RSU ($a_t = a_\mathrm{rsu}$). Therefore, the action space is designated as:
\begin{equation}\label{action}
A = V \cup \{a_\mathrm{cloud}\} \cup \{a_\mathrm{rsu}\}
\end{equation} 

\subsection{Reward Function}
The reward is a mathematical function that assigns a numerical value to each state-action pair. In this case, the reward function has been designed to provide a +5 reward for each completed task. Furthermore, the function deducts points equivalent to the total delay incurred during task completion, incentivizing the agent to complete tasks efficiently:
\begin{equation}\label{reward}
R(s,a) = \sum_{x \in U}^{} -d_\mathrm{total}^x + 5
\end{equation} 
where $U$ is the set of unaccounted tasks that have completed processing between the previous two states, and $d_\mathrm{total}^x$ specifies the total delay experienced by task $x$.

\section{Performance Evaluation}
This section details the simulation setup and explores the various results of the proposed method.
\subsection{Simulation Setup}
The simulated area is a 1000 by 1000 meter region modeled after downtown Salt Lake City (Figure 3) where each episode lasts for one minute. The maximum speed for vehicles is set at 48 km/h, consistent with urban district regulations. Details of the environment settings can be found in Table II. The three distinct scenarios considered, only differentiating in the number of vehicles, are as follows:
\begin{itemize}
\item {\emph{Scenario 1}: 5 client vehicles and 2 service vehicles.}
\item {\emph{Scenario 2}: 10 client vehicles and 4 service vehicles.}
\item {\emph{Scenario 3}: 20 client vehicles and 8 service vehicles.}
\end{itemize}

\begin{table}[!t]
\renewcommand{\arraystretch}{1.3}
\caption{Environment Settings}
\label{env_set}
\centering
\begin{tabular}{|c||c|}
    \hline
	Parameter & Value\\
	\hline
	$l \times w$ & 1000 (m) $\times$ 1000 (m)\\
	$n$ & 5, 10, 20\\
	$m$ & 2, 4, 8\\
	$\lambda_i$ & 3 (tasks per sec)\\
	$vel_i$ & $\sim U\{10,20,25,40\}$ (km/h) \\
	 $SZ_x$ & $\sim U\{20,40\}$ (Mb)\\
	 $B$ & 40 (MHz)\\
	 $P_T$ & 1 (W)\\
	 $G_T ,G_R$ & 5 (dBi)\\
	 $N_0$ & 174 (dBm/Hz)\\
	 $CPU_\mathrm{rsu}$ & 1$\times$18,375 (MIPS)\\
	 $CPU_\mathrm{cloud}$ & $\infty \times$100,000 (MIPS)\\
    \hline
\end{tabular}
\end{table}

RSU and service vehicle hardware are modeled based on specifications of real world processors. $CPU_j$ is uniformly sampled from the list at Table III, while $CPU_\mathrm{rsu}$ and $CPU_\mathrm{cloud}$ have constant values (Table II). An infinite amount of available hardware is assumed to exist in the cloud, this ensures that tasks delegated to the cloud will never encounter delays due to a lack of computational resources. However, before reaching the cloud, tasks must first be transmitted via the RSU. This occurs at a rate of 1 Gbps. Afterward, tasks navigate through the internet and experience various forms of propagation, transmission and queuing delays. The culmination of which is represented by a random value ranging from 0.05 to 0.2 seconds, capturing the variability of network delays.
\begin{table}[!t]
\renewcommand{\arraystretch}{1.3}
\caption{Processor Specifications}
\label{processor-specs}
\centering
\begin{tabular}{|c||c|}
    \hline
	Processor Name & Processor Speed (MIPS)\\
	\hline
	ARM Cortex A72 & 18,375\\
	AMD Phenom II X4 940 & 42,820\\
	ARM Cortex A73 & 71,120\\
    \hline
\end{tabular}
\end{table}  

The generated tasks in this study are based on a scaled-down version of the SPEC CPU 95 benchmark, a well-known standard for evaluating hardware performance. This benchmark consists of 18 tasks with varying computational complexities, ranging from 1.1 to 520 billion instructions. For a complete list of tasks and instruction counts, refer to \cite{phansalkar2005measuring}.

PPO was selected as the applied DRL algorithm. This method optimizes policies through a clipped objective function and has proven its effectiveness across a wide range of diverse tasks \cite{schulman2017proximal}.

\subsection{Delay Evaluation}
Three alternative offloading methods are considered for comparision:
\begin{itemize}
\item {\emph{Random offloading}: this basic method allocates each task to a randomly selected service node.}
\item {\emph{Cloud-only offloading}: offloads all incoming traffic to the cloud, ignoring the computational capabilities of the RSU and service vehicles.}
\item {\emph{Greedy offloading}: greedily selects the service node with the least congested transmission plus computational queue.}
\end{itemize}

Arguably, the most crucial performance metric of an offloading method is average task delay. Figure 5 plots this metric for the different offloading methods, with more details  available at Table IV. The RL algorithm consistently outperforms other techniques, with its superiority becoming more pronounced as the number of vehicles increases. Cloud-only offloading performs relatively efficiently with a small number of client cars (scenarios 1\&2), however, a noticeable dip in performance arises during the third scenario. The cause behind this inconsistent gap in is discussed in the following subsection.

\begin{figure}[!t]
\centering
\includegraphics[width=10cm]{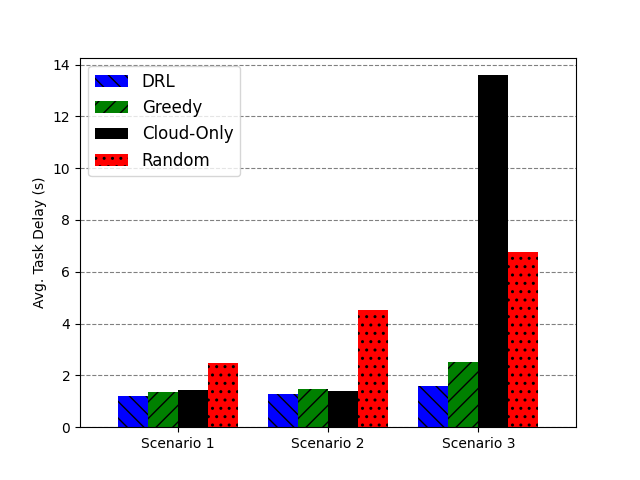}
\caption{Average task delay of various offloading methods; each column is the mean of 100 independent runs.}
\label{test_results}
\end{figure}

\subsection{Queue Length Evaluation}
Unlike the typically fixed computation and transmission delays, queuing delay in networking systems is a variable component that can fluctuate between tasks. The amount of this delay is influenced by the size of a queue, ranging from minimal when queues are empty to significant when they are full. Therefore, a sub-goal of any proper offloading algorithm should be controlling the buildup of queues across the network. Figure 6 displays the transmission queue size throughout an episode for all four methods.

\begin{table}[!t]
\renewcommand{\arraystretch}{1.3}
\caption{Average task delay with 95\% confidence intervals based on 100 samples.}
\label{avg-delay}
\centering
\begin{tabular}{|c|c|c|c|}
    	\hline
	\textbf{Offloading Method} & \textbf{Scenario 1} & \textbf{Scenario 2} & \textbf{Scenario 3}\\
	\hline
	RL & 1.22$\pm$0.11 & 1.27$\pm$0.08 & 1.59$\pm$0.05\\
	\hline
	Greedy & 1.35$\pm$0.11 & 1.48$\pm$0.08 & 2.53$\pm$0.12\\
	\hline
	Cloud-Only & 1.42$\pm$0.10 & 1.39$\pm$0.09 & 13.59$\pm$0.09\\
	\hline
	Random & 2.50$\pm$0.19 & 4.54$\pm$0.24 & 6.75$\pm$0.29\\
    	\hline
\end{tabular}
\end{table}   

The subpar performance of cloud-only offloading in scenario 3 can be explained by analyzing Traffic Intensity (TI) at the cloud transmission queue. This key metric is defined as the ratio of $aL/R$ where $a$, $L$ and $R$ represent the average packet arrival rate, average packet length and transmission rate, respectively. TI is widely used to estimate the extent of the queuing delay at a particular node. A TI value greater than 1 indicates that the average arrival rate (in bits) has exceeded the rate at which these bits can be transmitted from the queue. For cloud-based offloading in scenario 2, $TI = (3\times10)\times30/1000 = 0.9 < 1$, while in scenario 3,  $TI = (3\times20)\times30/1000 = 1.8 > 1$. This explains the basis of cloud-only offloading's performance drop as the number of vehicles, and thereby tasks, grows. Although increasing the transmission rate to the cloud (e.g. from 1Gbps to 10/100Gbps) can help alleviate this problem in the short term, the fact remains that in a real-life setting with ten's of thousands of vehicles, even this increased transmission power will fall short. This truth showcases the inefficiency and nonscalability of traditional cloud computing for a dynamic and high-load VFC environment. In comparison, offloading methods that utilize fog computing capabilities are inherently self-scalable (i.g., with more clients vehicles comes more service vehicles). Furthermore, intelligent offloading methods such as RL can learn to exploit fog-node offloading by choosing the most appropriate node based on proximity and hardware capabilities.

\begin{figure}[!t]
\centering
\includegraphics[width=10cm]{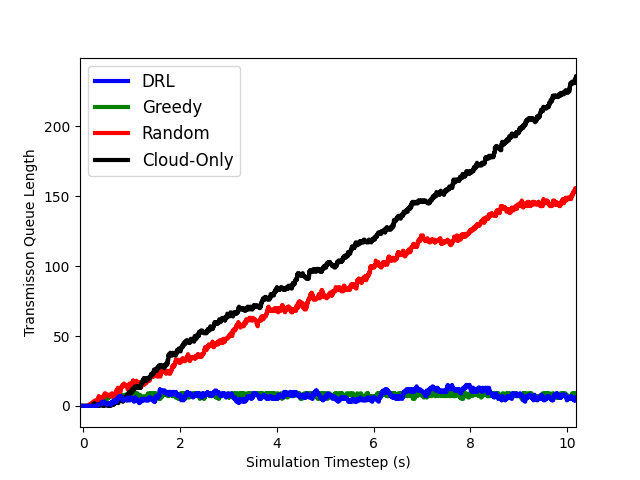}
\caption{Total length of all transmission queues throughout the first 10 seconds of simulation (Scenario 3).}
\label{queue_results}
\end{figure}

\section{Conclusion}
The objective of this research was investigating the effectiveness of reinforcement learning in task offloading within a vehicular fog computing environment. To this end, a realistic simulator making use of a novel movement model was designed. The results show that after training, the reinforcement learning algorithm consistently outperformed alternative offloading methods, significantly reducing queue congestion and ultimately average task delay. 

\bibliographystyle{plain} 
\bibliography{template} 

\end{document}